\documentclass[
 reprint,
superscriptaddress,
nofootinbib,
 amsmath,amssymb,
 aps,
prb,
]{revtex4-2}

\usepackage{siunitx}
\usepackage{graphicx}
\usepackage{dcolumn}
\usepackage{booktabs}
\usepackage{upgreek}

\usepackage{bm}

\begin{document}


\title{Effects of localized laser-induced heating in the photoluminescence of silicon-vacancy color centers in 4H-SiC}

\author{Adolfo Misiara}
 \affiliation{
 Department of Physics, Florida International University, Miami, FL 33199, USA
}%

\author{Stephen Revesz}
 \affiliation{
 Department of Physics, Florida International University, Miami, FL 33199, USA
}%

\author{Ibrahim Boulares}
 \affiliation{
U.S. Army Combat Capabilities Development Command, Army Research Laboratory, Adelphi, MD 20783, USA
}%

\author{Hebin Li}
 \email{hebinli@miami.edu}
\affiliation{
 Department of Physics, University of Miami, Coral Gables, FL 33124, USA
}
\affiliation{
 Department of Physics, Florida International University, Miami, FL 33199, USA
}%

\date{\today}

\begin{abstract}
Silicon vacancy ($\mathrm{V_{Si}}$) color centers in 4H-SiC are optically accessible through their zero-phonon line (ZPL) photoluminescence (PL), which is sensitive to the sample temperature. We report the effects of localized laser-induced heating in 4H-SiC by measuring the PL spectra of $\mathrm{V_{Si}}$ color centers. The effects of laser-induced heating manifest as the decrease in the peak height, redshift, and broadening of the ZPLs in the PL spectrum. The local temperature in the sample can be determined from the center energy of the ZPLs by using the Varshni equation. The sample temperature can be modeled as a system in contact with a thermal reservoir while being heated by a laser beam. This work highlights the importance of considering laser-induced heating in the optical properties of color centers in 4H-SiC and their potential applications. The result also suggests that the sharp and bright ZPLs of color centers can be used as local temperature probes in 4H-SiC devices. 
\end{abstract}

\maketitle


\section{\label{sec:Intro}Introduction}
The quantum technology revolution has propelled researchers to pursue new materials with innovative applications. Color centers are solid-state point defects with atom-like properties, making them a popular research topic for investigating their applicability in quantum sensing, quantum communication, and quantum computing technologies \cite{castelletto2020silicon, thiering2020colordiamond}. The fabrication of color centers has also benefited from advances towards more deterministic defect creation via controlled implantation in crystal lattices \cite{fuchs2015engineering, castelletto2020silicon, thiering2020colordiamond}. Silicon vacancy ($\mathrm{V_{Si}}$) color centers in 4H-SiC, in particular, enjoy a mature process for manufacturing high-purity 4H-SiC wafers \cite{maboudian2013advances, kimoto2014fundamentals} and color centers can be created through a variety of methods, including controlled ion, neutron, and electron irradiation \cite{castelletto2020silicon, kraus2017three, kasper2020influence}. Quantum applications based on $\mathrm{V_{Si}}$ color centers in 4H-SiC benefit from long electron spin-lattice relaxation times ($T_1$) \cite{widmann2015coherent, kraus2017three, simin2017locking, kasper2020influence} and electron spin-coherence times ($T_2$) \cite{yang2014electron, widmann2015coherent, carter2015spin, kraus2017three, simin2017locking, embley2017electron, nagy2018quantum, nagy2019high, brereton2020spin, lekavicius2022orders}, high quantum fidelities ($\mathcal{F}$) \cite{simin2017locking, nagy2019high}, and robust spin-photon interfaces \cite{soykal2016silicon, nagy2019high}. They can also be employed in high-precision magnetometry \cite{Simin2015High, kraus2014magnetic} and thermometry \cite{kraus2014magnetic}. 

These quantum applications employ the optical emission of the color centers' zero-phonon lines (ZPLs) for optical access. $\mathrm{V_{Si}}$ color centers in 4H-SiC exhibit sharp and robust ZPLs in the near-infrared region of the photoluminescence (PL) spectrum \cite{udvarhelyi2020vibronic}, known as V1 (1.438 eV), $\mathrm{V1^\prime}$ (1.443 eV), and V2 (1.352 eV). These ZPL emissions, along with properties indicative of quantum performance, such as $T_1$, $T_2$, and $\mathcal{F}$, have been shown to depend on sample temperature for $\mathrm{V_{Si}}$ color centers in 4H-SiC \cite{simin2017locking, embley2017electron}. The optical excitation of a laser can cause significant thermal effects and should be considered when studying the properties of these color centers. 

Here, we report the effects of localized laser-induced heating by probing the PL of $\mathrm{V_{Si}}$ color centers in 4H-SiC. The effects of laser-induced heating are manifested as the decrease in the peak height, redshift, and broadening of the V1, $\mathrm{V1^\prime}$, and V2 
ZPLs in the PL spectrum. These spectral changes have been associated with increasing temperature in semiconductors and color centers in previously published works \cite{castelletto2014siliconSM, nagy2018quantum}. Furthermore, similar changes in the PL caused by localized laser-induced heating have been observed in other materials such as $\mathrm{WS_2}$ monolayers \cite{Peng2021_LocalLaserHeatingWS2}, $\mathrm{ZnO}$ nanocrystals and nanostructures \cite{bergman2004photoluminescence, kurbanov2010uvZnOnanocrystal, yang2006photoluminescenceZnO}, $\mathrm{GaN}$ nanorods \cite{bergman2003impactGaN,bergman2004photoluminescence}, and silicon color centers in diamond \cite{gao2024local}. In this work, we apply the Varshni formula \cite{varshni1967temperature} to characterize the relationship between the center energy of the ZPLs and the sample temperature, while using a sufficiently low excitation power to ensure negligible heating. Using this relationship, we quantify the temperature variation induced by higher laser excitation powers and develop a mathematical model to fit the observed dynamics due to laser heating. The study highlights the importance of accounting for local heating by the excitation laser when investigating the optical properties of color centers in 4H-SiC. Moreover, expanding on this work, the narrow and bright ZPLs at low temperatures can allow the construction of local temperature probes and enable accurate localized temperature monitoring in SiC miniature optoelectronic devices. 

The rest of this article is organized as follows. Section \ref{sec:Details} describes the 4H-SiC sample studied and our PL setup. Section \ref{sec:Temporal} discusses the effects that laser power has over time in the PL spectrum of the ZPLs. It also shows that the temporal spectral changes caused by high excitation powers are not damaging or photo-bleaching the sample. Section \ref{sec:LowP} demonstrates that the spectral changes that occur with increasing temperature are the same qualitatively as the temporal spectral changes for high-power measurements. This section also characterizes the relationship between the center energy of the ZPLs and temperature using the Varshni formula \cite{varshni1967temperature}. Section \ref{sec:HighP} studies the localized laser-induced heating dependence on cryostat temperature and excitation power. Section \ref{sec:Model} uses the Varshni formula with the parameters obtained in Section \ref{sec:LowP} to calculate the change in the local temperature of the emitters on time. It also models the local temperature of the emitters as a system in contact with a heat reservoir being heated by laser light. Finally, Section \ref{sec:Conclusion} presents a summary of the work.

\begin{figure}[t]
    \centering
    \includegraphics[width=\columnwidth]{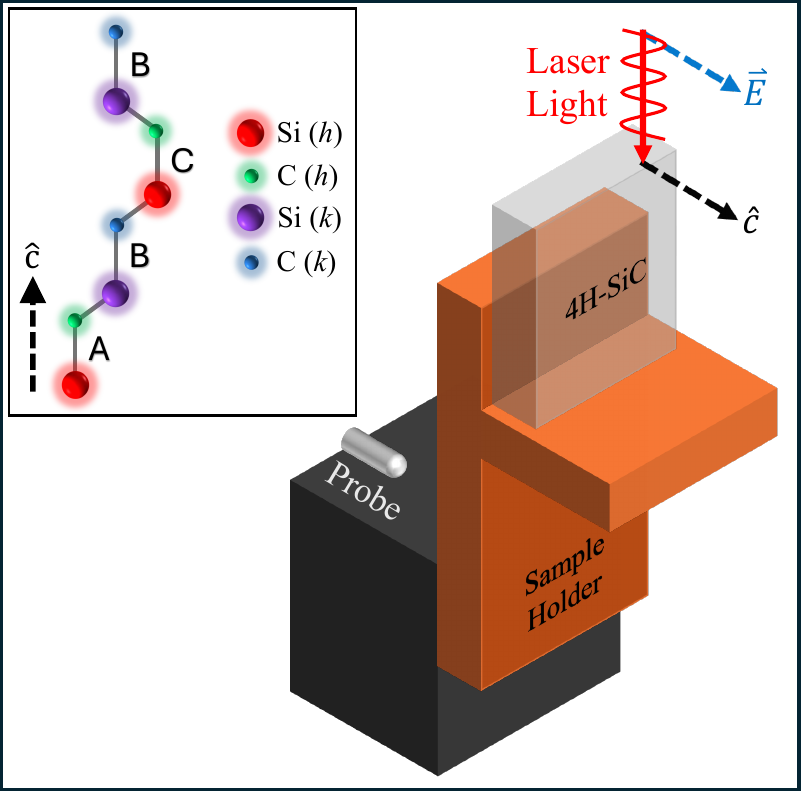}
    \caption{Schematic of the 4H-SiC sample thin the cryostat mounted on a sample holder in the side-view orientation. The sample is excited with laser light (red arrow) whose polarization (dashed blue arrow) is parallel to the c-axis (dashed black arrow). The temperature probe is mounted next to the sample holder. The inset shows the crystal lattice structure of 4H-SiC, displaying four Si-C bilayers in its unit cell stacked in sequence ABCB along the crystal c-axis \cite{kimoto2019sic}. Its crystal structure has two nonequivalent crystallographic sites, hexagonal ($h$) and cubic ($k$) \cite{udvarhelyi2020vibronic}. The legend of the inset shows how to identify the Si and C atoms in the $h-$ and $k-$sites within the unit cell.}
    \label{fig:SampleMount}
\end{figure}

\section{\label{sec:Details}Sample and photoluminescence experiment}

This section gives our sample details and describes the PL experimental setup. 

The sample (purchased from Wolfspeed) consists of a 20 $\mu$m-thick n-type 4H-SiC epilayer grown by chemical vapor deposition. It is nitrogen doped at a concentration of $10^{14}$ cm$^{-3}$ and its surface orientation is 4$^{\circ}$. The sample was irradiated by an electron fluence of $10^{18}$ $\mathrm{cm^{-2}}$ to create the $\mathrm{V_{Si}}$ color centers. A schematic of the 4H-SiC crystal lattice structure is included in the inset of Fig. \ref{fig:SampleMount}. It is a hexagonal crystal system with four Si-C bilayers in its unit cell stacked in sequence ABCB along the crystal c-axis \cite{kimoto2019sic}. The crystal structure has two nonequivalent crystallographic sites, hexagonal and cubic, called $h$- and $k$-sites, respectively \cite{magnusson2018excitation}. The legend of the inset in Fig. \ref{fig:SampleMount} shows how to identify the Si and C atoms in the $h$- and $k$-sites within the unit cell. A $h$-site $\mathrm{V_{Si}}$ gives rise to the V1 and $\mathrm{V1^\prime}$ ZPLs centered at 1.438 eV and 1.443 eV, respectively \cite{udvarhelyi2020vibronic}. Similarly, a $k$-site $\mathrm{V_{Si}}$ generates the V2 ZPL at 1.352 eV \cite{udvarhelyi2020vibronic}. To the best of our knowledge, the $\mathrm{V2^\prime}$ ZPL proposed to be 22 meV higher in energy than V2 \cite{udvarhelyi2020vibronic} has not been observed to date. 

The sample was housed in a Montana Instruments closed-cycle liquid helium cryostat capable of cooling the sample down to 4 K. The cryostat maintains a vacuum pressure lower than $\mathrm{10^{-3}}$ torr. The sample was mounted on a custom-built sample holder, as shown in Fig. \ref{fig:SampleMount}, in the so-called side-view orientation so that the cleaved edge of the sample is optically accessible. The sample was excited by a laser beam orthogonal to the c-axis of the 4H-SiC crystalline structure. The polarization of the laser was rotated using a half-wave plate so that the electric field was parallel to the c-axis, $\vec{E}\parallel\hat{c}$. All measurements were performed in the side-view orientation of the sample (which favors V1 and V2 emission over $\mathrm{V1^\prime}$ \cite{udvarhelyi2020vibronic}) with $\vec{E}\parallel\hat{c}$. As shown in Fig. \ref{fig:SampleMount}, the temperature probe used to measure the cryostat temperature is not in direct contact with the sample. 

\begin{figure}[htb]
    \centering
    \includegraphics[width=\columnwidth]{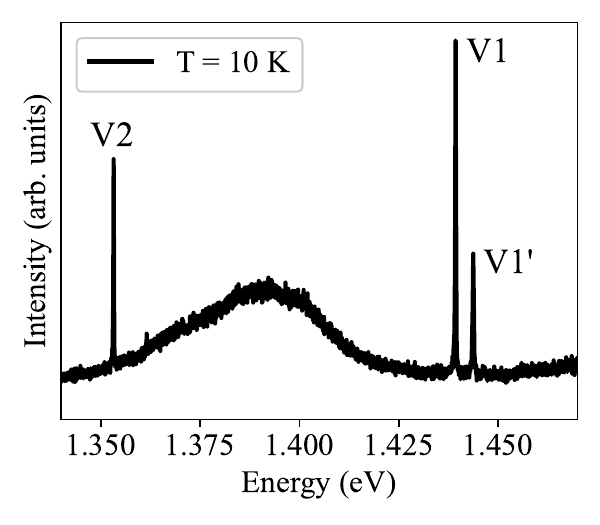}
    \caption{PL spectrum of $\mathrm{V_{Si}}$ color centers in 4H-SiC at a cryostat temperature of 10 K for a power of 0.03 mW. The predominant V2, V1, and $\mathrm{V1^\prime}$ ZPLs are labeled from left to right. The PSB can be observed between the V1 and V2 ZPLs.}
    \label{fig:PL_FullSpec}
\end{figure}

The PL experiment was performed with a PL microscope setup. The sample was excited by a continuous-wave external-cavity diode laser with a wavelength of 780 nm (1.59 eV). The excitation laser was focused onto the sample with a long-working distance 50X Mitutoyo plan apochromat objective (NA = 0.55). The same objective was used to collect the PL signal, which was then filtered by a notch filter (centered at 785 nm with a bandwidth of 33 nm) to remove reflected and scattered light from the 780 nm excitation laser. A plano-convex lens with a focal lens of 50 mm focused the filtered PL signal into a multi-mode optical fiber whose output fed into a half-meter spectrometer (Horiba iHR-550). The spectrometer used a 1200 gr/mm grating integrated with a charged-couple device (CCD) camera. The entrance slit size of the spectrometer was 50 $\upmu$m. The spectral resolution of the spectrometer was measured to be 0.123 meV using the emission line at 877.67 nm from a calibration Krypton lamp. A typical PL spectrum of $\mathrm{V_{Si}}$ color centers in 4H-SiC is shown in Fig. \ref{fig:PL_FullSpec}. This spectrum was taken at a cryostat temperature of 10 K for an excitation power of 0.03 mW. The predominant V2, V1, and $\mathrm{V1^\prime}$ ZPLs are shown, labeled from left to right. The spectrum also features a phonon-sideband (PSB) between V1 and V2.



\section{\label{sec:Temporal} Effects of laser heating on photoluminescence spectra}

\begin{figure}[htb]
    \centering
    \includegraphics[width=\columnwidth]{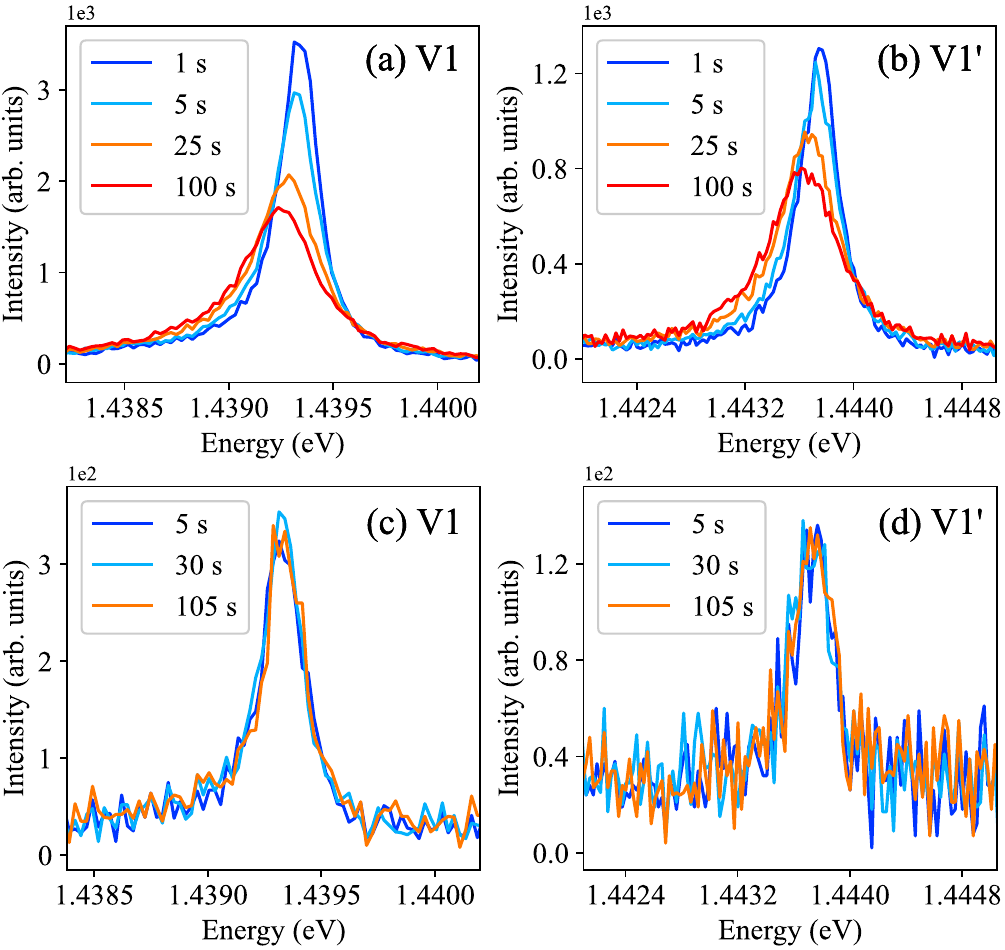}

    \caption{Timed PL measurements conducted at a cryostat temperature of 10 K. The legends indicate how long the laser illuminated the sample when the spectrum was taken. The (a) V1 and (b) $\mathrm{V1^{\prime}}$ ZPLs decrease in peak height, redshift, and broaden over time when excited with a power of 2 mW. The spectral characteristics of (c) V1 and (d) $\mathrm{V1^{\prime}}$ ZPLs do not change significantly over time when excited with a power of 0.03 mW.}
   
    \label{fig:10K_Decay}
\end{figure}

To study the effects of laser-induced heating on the ZPL PL spectra, we measured PL spectra at different times after the excitation laser started illuminating the sample. During the experiment, the laser was unblocked, and the PL was collected in timed PL measurements, which continued for a total time of $\sim$ 135 s. A set of measurements, lasting 135 s in total, had a fixed collection time for the individual PL measurements that ranged from 0.25 s to 2 s. The collection time was selected according to the lowest signal-to-noise ratio (higher collection times were used for weaker signals) achieved for that set of measurements, which depended on the cryostat temperature and excitation power. All PL spectra were normalized by dividing each spectrum by its collection time to account for the different collection times. The timed PL measurements were done with the excitation laser at various powers. High-power measurements were conducted at discrete power levels ranging from 0.5 mW to 2.0 mW in 0.5 mW increments, corresponding to intensities ranging from 62.5 W/$\mathrm{mm^{2}}$ to 250 W/$\mathrm{mm^{2}}$ for a beam size of $\sim{8} $ \SI{}{\micro\meter\squared}. As an example, the PL spectra at times $t=1, 5, 25,$ and $100$ s obtained with an excitation laser power of 2 mW are shown in Fig. \ref{fig:10K_Decay} for both (a) V1 and (b) $\mathrm{V1^\prime}$ at a cryostat temperature of 10 K. These PL measurements had a collection time of 0.25 s. At these excitation powers, timed PL spectra showed a decrease in the peak height, a redshift of the center energy, and a broadening of the linewidth of the ZPLs over time while the excitation laser illuminated the sample. 

In comparison, we also performed low-power PL measurements with an excitation laser power of 0.03 mW, corresponding to an intensity of 3.75 W/$\mathrm{mm^{2}}$. The examples of timed PL spectra for this excitation power are shown in Fig. \ref{fig:10K_Decay}(c) and (d) for the V1 and $\mathrm{V1^\prime}$ ZPLs, respectively. These PL measurements had a collection time of 1.5 s. At this excitation power, the spectral characteristics of the PL did not significantly change over time, suggesting that laser heating is negligible at this power and does not affect the sample temperature. All measurements in Fig. \ref{fig:10K_Decay} were taken at a cryostat temperature of 10 K. The same behavior with high and low laser powers were observed for the timed PL measurements of the V2 ZPL. The results for V2 are shown in Appendix \ref{Apx:PL}. 

\begin{figure}[htb]
    \centering
    \includegraphics[width=\columnwidth]{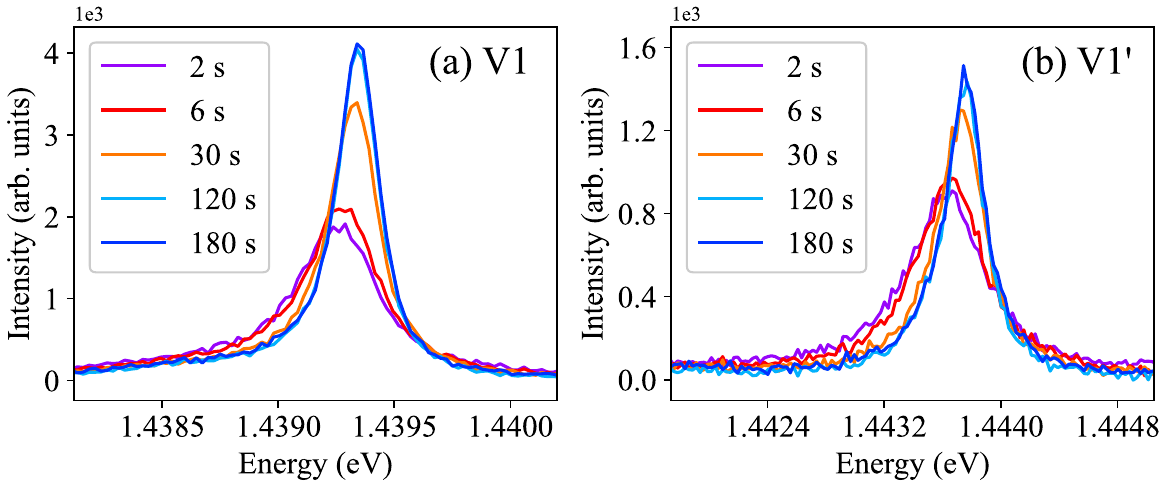}

    \caption{Recovery of the ZPL PL at a cryostat temperature of 10 K. The sample was continuously excited with 2 mW of power for 180 s before the signal was allowed to recover (the laser was blocked) for a selected period of time. After the recovery time passed, the PL was measured again right after the laser was unblocked. The legends indicate the recovery time, $t_R$, in seconds before the spectrum was taken. The (a) V1 and (b) $\mathrm{V1^{\prime}}$ ZPLs increase in peak height, blueshift, and narrow as the recovery time increases.}
   
    \label{fig:10K_Recovery}
\end{figure}

The changes to the spectral characteristics of the high-power PL measurements were not permanent. Preventing the laser from illuminating the sample for longer than 120 s allowed the PL signal to recover and the timed PL measurements to be repeated. To study the recovery of the PL signal, we excited the sample for 180 s before proceeding to block the laser for a predetermined recovery time, $t_R$. After the chosen recovery time passed, the laser was unblocked, and the first PL spectrum was taken as the recovered signal. Fig. \ref{fig:10K_Recovery} shows the results of this experiment conducted for an excitation power of 2 mW at a cryostat temperature of 10 K for the (a) V1 and (b) $\mathrm{V1^\prime}$ ZPLs. The legends of the figure indicate $t_R$ for the respective spectrum. It shows that as $t_R$ increases, the ZPLs increase in peak height, blueshift, and narrow. The experiment demonstrates that the PL signal recovers its original PL spectrum over time after blocking the excitation laser. The depletion of the PL signal would be permanent if the excitation laser damages or photo-bleaches the sample. In the following sections, we show that the temporal behavior of the ZPL PL for high excitation powers can be attributed to the change in sample temperature due to localized laser-induced heating.

\section{\label{sec:LowP}Temperature dependence of low-power PL}

In this section, we performed a temperature dependence measurement of the ZPL PL at low power to show experimentally that increasing temperature leads to a decreased peak height, redshift, and broadening of the ZPLs. The PL temperature dependence was conducted with a low power, 0.03 mW, to avoid the effects of laser-induced heating. Using the cryostat temperature as the sample temperature for the low-power measurements allowed us to quantitatively characterize the relationship between the center energy of the ZPLs and temperature.

\begin{figure}[!]
    \centering
    \includegraphics[width=\columnwidth]{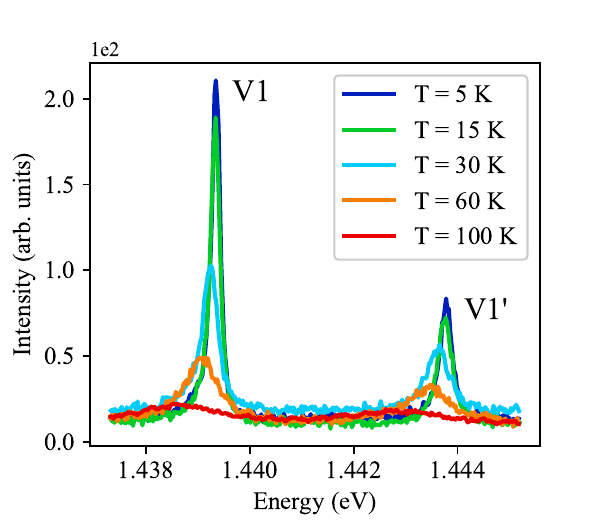}
    \caption{PL spectra taken with an excitation power of 0.03 mW at cryostat temperatures 5 K, 15 K, 30 K, 60 K, and 100 K. The V1 and $\mathrm{V1^\prime}$ ZPLs decrease in peak height, redshift, and broaden with increasing temperature.}
    \label{fig:TempDepPL}
\end{figure}

The low-power PL measurements were taken at cryostat temperatures 5 K, 10 K, 15 K, 20 K, 30 K, 40 K, 60 K, 80 K, 100 K, and 125 K. PL spectra of the V1 and V1' ZPLs at select temperatures are displayed in Fig. \ref{fig:TempDepPL}. The legend of the figure indicates the cryostat temperature at which the spectrum was taken. A decrease in the peak height, redshift of the center energy, and broadening of the linewidth with increasing temperature can be seen in the figure. The behavior of the ZPL PL with increasing temperature, seen in Fig. \ref{fig:TempDepPL}, resembles the PL behavior over time taken with high powers, seen in Fig. \ref{fig:10K_Decay}(a) and (b). In both cases, the ZPLs decrease in peak height, redshift, and broaden. This observation supports the hypothesis that the temporal spectral changes for high excitation powers could result from the change in sample temperature due to localized laser-induced heating.

The low-power temperature dependence of the ZPL PL can be used to characterize the effects of sample temperature in the PL spectral features. In particular, we studied the temperature dependence of the center energy of the ZPLs. The center energies can be extracted by fitting the PL spectra with the Lorentzian lineshape with a linear baseline. The linear baseline was added to account for the overlap between the wings of the PSB and the ZPLs (Fig. \ref{fig:PL_FullSpec}). Appendix \ref{Apx:Fitting} includes a description and examples of the fitting. The peak height and linewidth of the ZPLs can also be extracted from the fitting. During the experiment, three PL measurements were taken at each temperature, and the center energy was determined by fitting each spectrum. The center energy mean, $\bar{E}_0$, and sample standard deviation, $S$, were calculated from the three measurements at each temperature.

\begin{figure}[htb]
    \centering
    \includegraphics[width=\columnwidth]{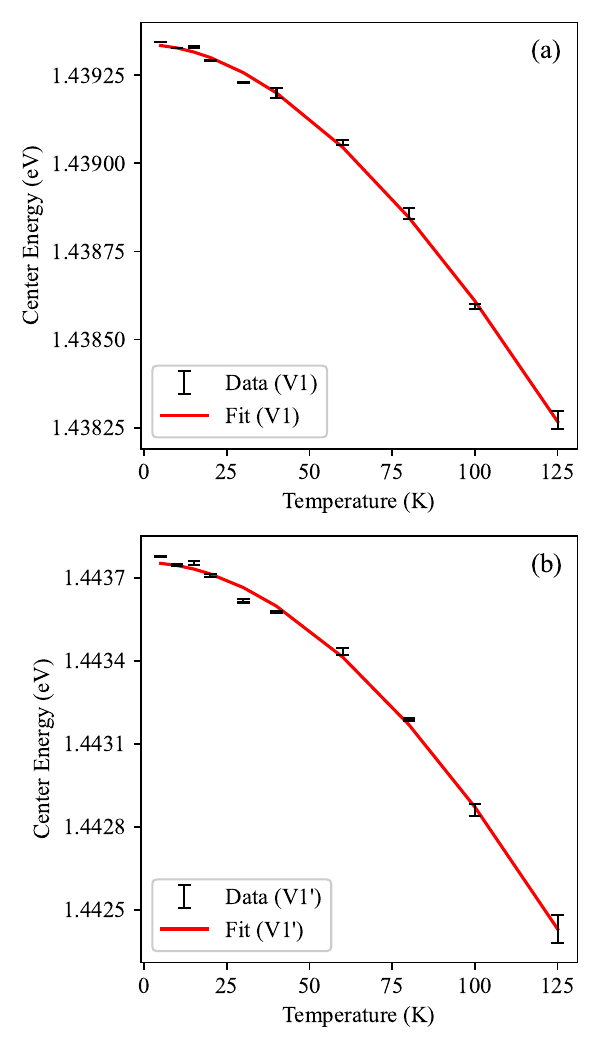}

    \caption{Center energy versus temperature data for (a) V1 and (b) $\mathrm{V1^\prime}$ measured for an excitation power of 0.03 mW. The Varshni equation was fitted to the data. The fits are displayed as red lines.}
   
    \label{fig:VarV1}
\end{figure}

The previous results, $\bar{E}_0$ for (a) V1 and (b) $\mathrm{V1^\prime}$ were plotted versus temperature in Fig. \ref{fig:VarV1} and the error bars of the figure were made from $S$. To establish a quantitative relationship, these results were fitted with the Varshni equation, which has been used to describe the band gap energy temperature dependence in semiconductor materials \cite{varshni1967temperature}. 
The Varshni equation is 
\begin{equation}\label{eq:Varshni}
    \epsilon = \epsilon_0 - \frac{\alpha T^2}{\beta + T},
\end{equation}
\noindent where $\epsilon$ is the band gap energy, $\epsilon_0$ is the band gap energy at absolute zero, and $\alpha$ and $\beta$ are parameters dependent on the material. Under the assumption that changes to the band gap energy alter the spacing of all energy levels within it uniformly, we use the Varshni equation to model the center energy of V1 and $\mathrm{V1^\prime}$ emission. Previous studies on temperature-dependent line shifts of defect centers in semiconductors have introduced models based on a modified Varshni equation \cite{Li2017} and Davies' formalism \cite{Davies1974}, the latter attributing ZPL shifts to lattice thermal expansion and electron-phonon interactions \cite{Revesz2024}. The fits displayed as the red lines in Fig. \ref{fig:VarV1} show that the Varshni formula is an appropriate mathematical expression to describe the change of the center energy with temperature. The fitting parameters of the Varshni fit for V1 and $\mathrm{V1^\prime}$ are shown in Table \ref{table:Varshni}. 

\begin{table}[htb]
\centering
\renewcommand{\arraystretch}{1.5} 
\setlength{\tabcolsep}{12pt} 
\begin{tabular}{|c|c|c|c|}
\hline
ZPL Name & $\epsilon_0$ (eV) & $\alpha$ $\mathrm{(\upmu eV/K)}$ & $\beta$ (K) \\
\hline
V1 & 1.4393 & 29 & 300 \\
\hline
$\mathrm{V1^\prime}$ & 1.4438 & 52 & 490 \\
\hline
V2 & 1.3532 & 27 & 380 \\
\hline
\end{tabular}
\caption{Fitting parameters of the Varshni fit for the V1, $\mathrm{V1^\prime}$, and V2 ZPLs.}
\label{table:Varshni}
\end{table}

A similar analysis for V2 is included in Appendix \ref{Apx:LowP}. Like for V1 and $\mathrm{V1^\prime}$, the low-power temperature dependence of V2 shows that increasing temperature leads to decreasing peak height, redshifting, and broadening of the V2 ZPL (Fig. \ref{fig:V2TempDep}). For completeness, the V2 ZPL was fitted with the Lorentzian lineshape with a linear baseline (examples of fitting for V2 are shown in Fig. \ref{fig:FittingV2}) as described in Appendix \ref{Apx:Fitting}, and its center energy was extracted. The Varshni formula was then used to fit the center energy mean versus temperature data (Fig. \ref{fig:VarV2}), similarly to V1 and $\mathrm{V1^\prime}$. The Varshni fitting parameters for V2 are displayed in Table \ref{table:Varshni}. 

\section{\label{sec:HighP}Temperature dependence of high-power PL} 

\begin{figure*}[!]
    \centering
    \includegraphics[width=\textwidth]{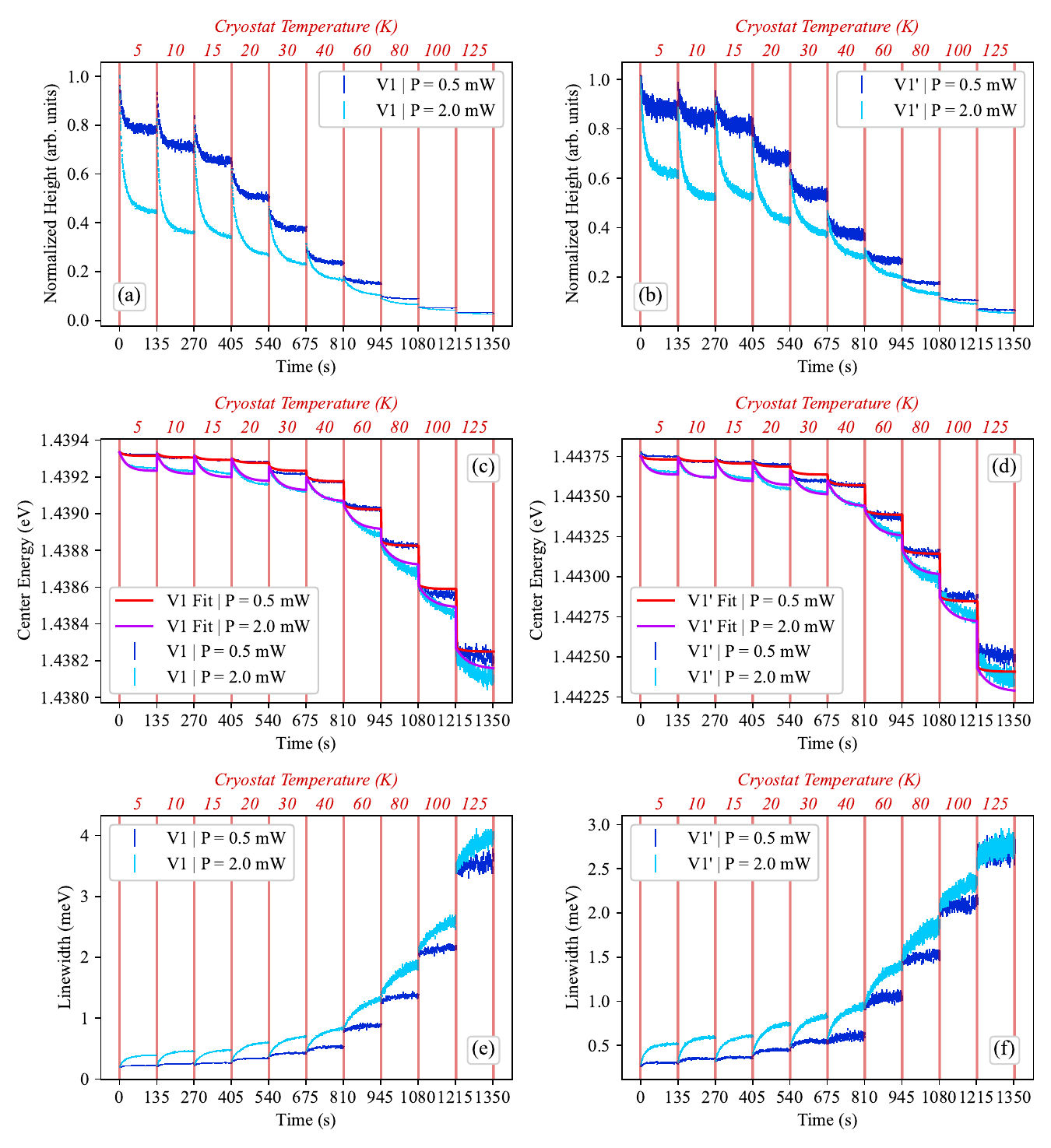}

    \caption{Temperature dependence of the spectral characteristics of V1 and $\mathrm{V1^\prime}$ over time for excitation powers of 0.5 mW (dark blue vertical bars) and 2.0 mW (light blue vertical bars). The normalized peak height of (a) V1 and (b) $\mathrm{V1^\prime}$, the center energy of (c) V1 and (d) $\mathrm{V1^\prime}$, and the linewidth of (e) V1 and (f) $\mathrm{V1^\prime}$ were extracted from fitting and graphed as a function of time. The data from a series of timed PL measurements at different cryostat temperatures are patched together so that the lower x-axis is composed of multiples of 135 s. The cryostat temperature of each series of timed PL measurements is displayed in the upper x-axis, and dull red lines separate timed PL measurements at different cryostat temperatures. 
    The (c) V1 and (d) $\mathrm{V1^\prime}$ center energies were fitted simultaneously (for measurements conducted at the same power) by finding the numerical solution of Eq. (\ref{eq:Model}) for the temperature and using Eq. (\ref{eq:Varshni}) to calculate the (a) V1 and (b) $\mathrm{V1^\prime}$ center energies. The result from fitting the (c) V1 and (d) $\mathrm{V1^\prime}$ center energies acquired using a power of 0.5 (2.0) mW is displayed as the red (purple) line.}
    \label{fig:Par_HighP}
\end{figure*}

To confirm that the PL spectral changes over time, as shown in Fig. \ref{fig:10K_Decay}(a) and (b), are due to laser-induced temperature changes in the sample, we repeated the high-power timed PL measurements at various cryostat temperatures. These results also aided in studying the effects of excitation power and cryostat temperature on the laser heating process.



The timed PL measurements at various cryostat temperatures were performed using excitation laser powers 0.5 mW, 1.0 mW, 1.5 mW, and 2.0 mW, corresponding to intensities ranging from 62.5 W/$\mathrm{mm^{2}}$ to 250 W/$\mathrm{mm^{2}}$. At each cryostat temperature, a series of timed PL spectra were obtained while the excitation laser illuminated the sample for up to 135 s. The measurements were conducted at cryostat temperatures 5 K, 10 K, 15 K, 20 K, 30 K, 40 K, 60 K, 80 K, 100 K, and 125 K. Fitting the obtained PL spectra with the Lorentzian lineshape plus a linear baseline (as described in Appendix \ref{Apx:Fitting}), we extracted the peak height, center energy, and linewidth of the ZPLs. The results are shown in Fig. \ref{fig:Par_HighP} for excitation powers 0.5 mW and 2.0 mW. The figure plots the extracted peak height, center energy, and linewidth as a function of the waiting time $t$ ($0 - 135$ s). The data from the series of timed PL measurements at different cryostat temperatures are patched together so that the lower x-axis is composed of multiples of 135 s, indicating the end of each series of timed PL measurements. The cryostat temperature of each series of timed PL measurements is displayed in the upper x-axis, and dull red lines separate timed PL measurements at different cryostat temperatures. The normalized peak height of the V1 and $\mathrm{V1^\prime}$ ZPLs over time at different cryostat temperatures is shown in (a) and (b), respectively. The V1 and $\mathrm{V1^\prime}$ peak heights of the first timed PL spectrum at a cryostat temperature of 5 K are normalized to one, and the same scaling factors were used to normalize the peak heights in all other PL spectra obtained with the same power. The center energy of the V1 and $\mathrm{V1^\prime}$ ZPLs is shown in (c) and (d), respectively, and the linewidth of the V1 and $\mathrm{V1^\prime}$ ZPLs is shown in (e) and (f), respectively. Each data point is marked by a vertical bar, with its length representing the fitting uncertainty. Larger uncertainties are observed at higher temperatures due to lower signal-to-noise ratios. Due to the weaker PL signals at higher temperatures, we also had to increase the collection time per PL measurement for higher temperatures, which decreased the time resolution (collection times ranged from 0.25 s to 2 s).

\begin{figure*}[hbt]
    \centering
    \includegraphics[width=\textwidth]{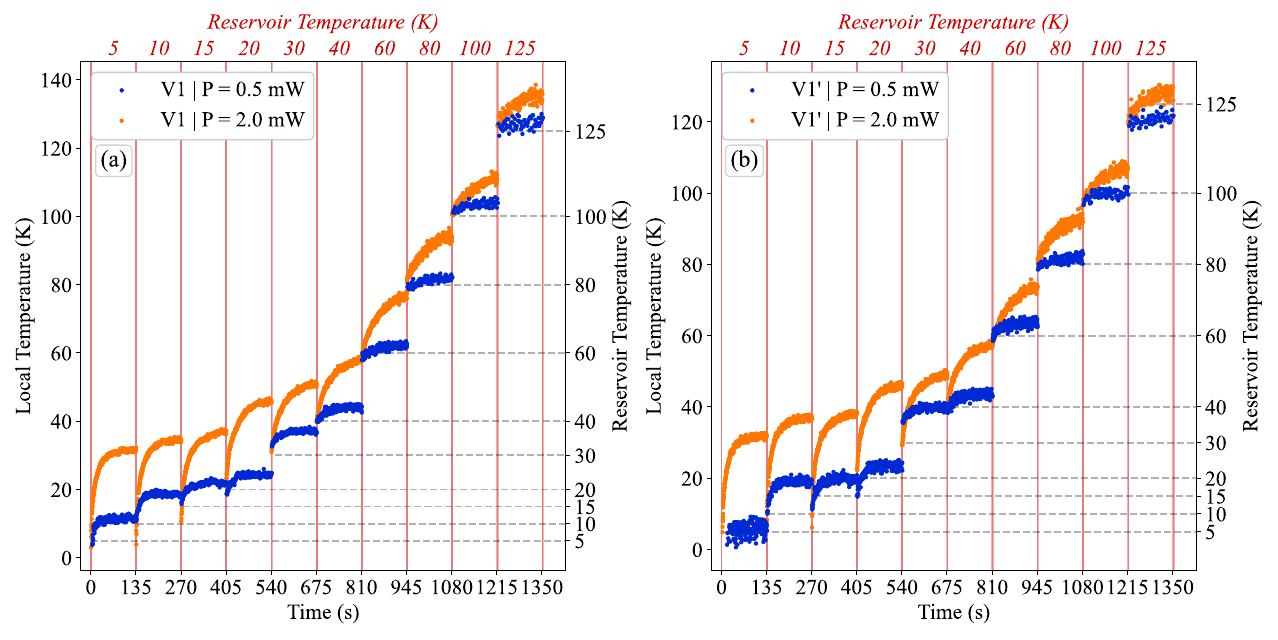}

    \caption{Local temperature of the emitters calculated from the center energy of the (a) V1 and (b) $\mathrm{V1^\prime}$ ZPLs using the Varshni formula. The redshift of the ZPLs was measured for 135 s at different reservoir temperatures and used to calculate the local temperature. The PL measurements were conducted with powers 0.5 mW (blue dots) and 2.0 mW (orange dots). The data at different reservoir temperatures are patched together so that the lower x-axis is composed of multiples of 135 s. The reservoir temperature of each heating process is displayed in the upper x-axis, and dull red lines separate heating processes at different reservoir temperatures. Dashed gray lines extending from the y-axis on the right indicate that the reservoir temperature is close to the initial local temperature of its respective heating process.}
   
    \label{fig:Tem_HighP}
\end{figure*}

In Fig. \ref{fig:Par_HighP}, it can be observed that the normalized peak height of the (a) V1 and (b) $\mathrm{V1^\prime}$ ZPLs decreases over time for all the cryostat temperatures, the center energy of (c) V1 and (d) $\mathrm{V1^\prime}$ ZPLs redshifts over time for all the cryostat temperatures, and the linewidth of (e) V1 and (f) $\mathrm{V1^\prime}$ ZPLs broadens over time for all the cryostat temperatures. The total changes in these parameters from their initial values to the values at the end of 135 s are smaller for higher cryostat temperatures. If these changes are induced by localized laser-induced heating, it suggests that the laser has a smaller heating effect for higher temperatures. It can also be observed that the total changes in these parameters are greater for higher powers at all cryostat temperatures. This occurs because a higher excitation power heats the sample more, leading to a higher final local temperature and, hence, a greater total change in the peak height, center energy, and linewidth. 

The previous remarks are further supported by an analysis of the initial values of the parameters at the beginning of the excitation for both powers. At all cryostat temperatures, the initial values ($t \approx 0$ s) of the normalized peak height, center energy, and linewidth are similar for the two powers. Simply put, the first normalized ZPL spectrum (taken at the beginning of the excitation) is always the same, independent of power, for all cryostat temperatures. This occurs because the first PL spectrum was collected only a brief amount of time after the excitation began, so the local temperature did not have enough time to change significantly. Therefore, the first timed PL spectra (with the peak height normalization described above) are almost identical for both powers. Only over time does the laser-induced heating affect the PL spectra and cause variations in the spectral features (peak height, center energy, and linewidth) until reaching an ``equilibrium" after about 120 s. The higher the excitation laser power, the more significant these variations are. 

Another observation that can be made from Fig. \ref{fig:Par_HighP} is that the total changes in the parameters from their initial values under an excitation power of 2 mW at a cryostat temperature of 5 K are equivalent to the changes that would occur due to a cryostat temperature increase of 25 K. In the data for the excitation power of 2 mW, the final equilibrium values (at $t=135$ s) of peak height, center energy, and linewidth at the cryostat temperature of 5 K are the same as the initial values (at $t=0$ s) of these parameters at 30 K. This suggests that 2 mW of power alters the local temperature in the sample from 5 K to 30 K over the course of 120 s and stabilizes at about 30 K afterward.


It is important to note that decreases in the peak height over time of the ZPL PL from divacancy defects (VV) in 4H-SiC have been observed by Magnusson, et. al. \cite{magnusson2018excitation}. In their work, the decrease in the intensity of the ZPLs has been attributed to population losses of the VV bright state ($\mathrm{VV^0}$) due to charge dynamics caused by the excitation source energy \cite{magnusson2018excitation, wolfowicz2017optical}. An excitation energy of 1.59 eV, like the one used in this work, has been shown to lead to charge stabilization of the $\mathrm{V_{Si}}$ bright state ($\mathrm{V_{Si}^{-}}$) and not to cause a depletion of the PL signal from $\mathrm{V_{Si}}$ color centers \cite{wolfowicz2017optical}. Hence, charge dynamics do not contribute to the decrease in the peak height over time seen in this work. 

\section{\label{sec:Model}Modeling of heating effects}

In this section, we use the Varshni equation to extract the local temperature variations induced by high-power excitation and model these changes by considering the laser-induced heating of a system in contact with a thermal reservoir. 

Using the parameters from the Varshni fitting described in Section \ref{sec:LowP}, the local temperature of the emitters can be calculated from the center energy as 

\begin{equation}\label{eq:Temp}
    T = \frac{\sqrt{(\epsilon-\epsilon_0)^2-4\alpha\beta(\epsilon-\epsilon_0)}-(\epsilon-\epsilon_0)}{2\alpha}.
\end{equation}
Using this equation, the center energies shown in Fig. \ref{fig:Par_HighP}(c) and (d) can be converted into the local temperatures of the emitters. The local temperatures extracted from the data of Fig. \ref{fig:Par_HighP}(c) and (d) are displayed in Fig. \ref{fig:Tem_HighP}(a) and (b), respectively, for excitation powers 0.5 mW and 2.0 mW. The following discussion refers to the cryostat temperature as the reservoir temperature $T_R$. The local temperature was found over time in a period of 135 s at reservoir temperatures: 5 K, 10 K, 15 K, 20 K, 30 K, 40 K, 60 K, 80 K, 100 K, and 125 K. The data at different reservoir temperatures are patched together so that the lower x-axis is composed of multiples of 135 s, indicating the end of each heating process. The reservoir temperature of each heating process is displayed in the upper x-axis and dull red lines separate heating processes at different reservoir temperatures. Dashed gray lines extending from the y-axis on the right indicate that the reservoir temperature is close to the initial local temperature of its respective heating process.

In Fig. \ref{fig:Tem_HighP}, it can be observed that the local temperature of the emitters increases over time upon laser excitation at all reservoir temperatures. Before excitation, the local temperature is close to the reservoir temperature. Upon excitation, the emitters are heated over time due to laser-induced heating and reach a thermal equilibrium after about 120 s. Since a higher power heats up the sample more, a higher final local temperature is achieved for 2 mW than 0.5 mW of excitation power. Also, the total changes in the local temperature due to the laser-induced heating are smaller for higher reservoir temperatures. These observations are consistent with the conclusions from Section \ref{sec:HighP}. The local temperatures found from the V1 ZPL data are similar to those found from the $\mathrm{V1^\prime}$ ZPL data under the same experimental conditions (excitation power, reservoir temperature, and waiting time). This is expected since the V1 and $\mathrm{V1^\prime}$ emitters should experience the same local temperature. 

\begin{figure}[!]
    \centering
    \includegraphics[width=\columnwidth]{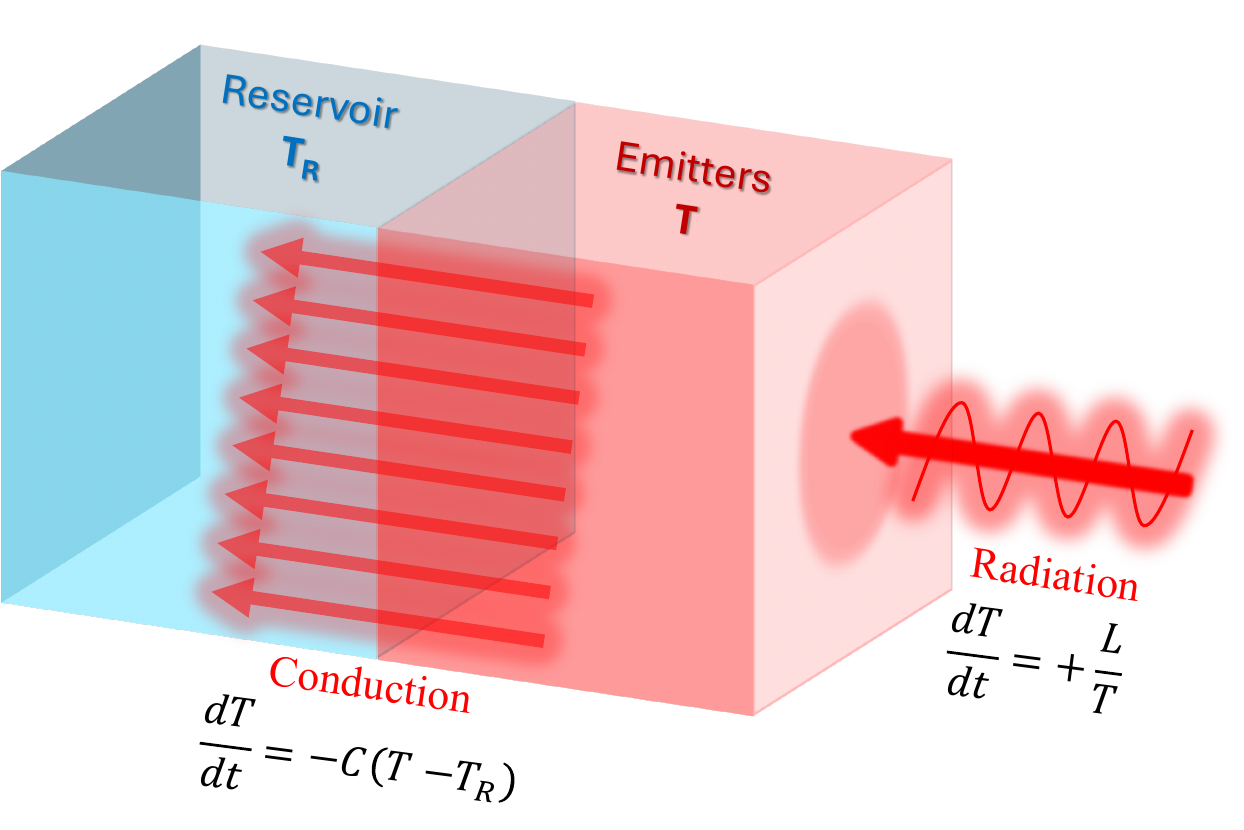}

    \caption{Schematic of the heating model. Emitters at temperature $T$ are heated by the excitation laser at rate $L/T$ and dissipate heat due to conduction to a thermal reservoir at temperature $T_R$ at rate $C(T-T_R)$. }
   
    \label{fig:Model}
\end{figure}


   

Considering the emitters as a system that is being heated by the excitation laser and is in direct contact with a thermal reservoir at temperature $T_R$, we can model the temperature change over time with the following equation
\begin{equation}\label{eq:Model}
    \frac{d{T}}{d{t}} = \frac{L}{T} - C(T-T_R ).
\end{equation}
\noindent A schematic depiction of this model is shown in Fig. \ref{fig:Model}. The system loses heat over time to the reservoir due to conduction and gains heat due to the radiation of the excitation laser. The rate of conduction is proportional to the difference in temperature between the system and the thermal reservoir by a factor $C$. The heating rate due to laser radiation is given by a constant $L$, which should be proportional to the laser power. We divide this term by the temperature, $T$, phenomenologically, after observing that greater temperature increases due to laser heating are produced for lower temperatures. Since the sample is in a vacuum, the heat exchange due to convection is negligible. We also assume that the heat lost due to radiative emission is insignificant. 

Solving the differential equation, Eq. (\ref{eq:Model}), numerically and converting temperature to center energy using the Varshni formula, Eq. (\ref{eq:Varshni}), with the obtained fitting parameters in Section \ref{sec:LowP}, we can fit the plots of Fig. \ref{fig:Par_HighP}(c) and (d). The center energy data of V1 and $\mathrm{V1^\prime}$ taken with the same laser power were fit simultaneously since the parameters $L$ and $C$ in Eq. (\ref{eq:Model}) must be the same for both types of emitter. The initial condition for solving Eq. (\ref{eq:Model}) is that the sample is at thermal equilibrium with the reservoir before excitation, so the initial temperature is $ T(t=0) = T_R$. The center energy data at different reservoir temperatures were also fit simultaneously while only changing the initial condition in accordance with the reservoir temperature. The fitting results for the (c) V1 and (d) $\mathrm{V1^{\prime}}$ data obtained with an excitation power of 0.5 (2.0) mW are displayed as the red (purple) line in Fig. \ref{fig:Par_HighP}. The model describes the data more appropriately at lower temperatures and fits the V1 data better than the $\mathrm{V1^\prime}$ data. This can be explained by the lower signal-to-noise ratio of the ZPLs at higher temperatures (Fig. \ref{fig:TempDepPL}) and the lower signal-to-noise of the $\mathrm{V1^\prime}$ ZPL compared to the V1 ZPL (Fig. \ref{fig:PL_FullSpec}). Also, the overlap between the V1 and $\mathrm{V1^\prime}$ ZPLs increases with temperature due to broadening, which introduces more uncertainty in determining the center energy of the ZPLs at higher temperatures.


\begin{table}[h]
\centering
\renewcommand{\arraystretch}{1.5} 
\setlength{\tabcolsep}{12pt} 
\begin{tabular}{|c|c|c|}
\hline
Power (mW) & L ($\mathrm{K^2/s}$) & C (mHz) \\
\hline
0.5 & $5.58 \pm 0.44$ & $36.0 \pm 3.4$ \\
\hline
1.0 & $11.31 \pm 0.33$ & $30.5 \pm 1.1$ \\
\hline
1.5 & $17.71 \pm 0.29$ & $27.2 \pm 0.6$ \\
\hline
2.0 & $21.14 \pm 0.22$ & $20.7 \pm 0.3$ \\
\hline
\end{tabular}
\caption{Fitting parameters of the temperature model for various powers.}
\label{table:Model}
\end{table}

Similar fits were done to data obtained with excitation powers 1.0 mW and 1.5 mW. The fitting parameters $L$ and $C$ obtained for powers 0.5 mW, 1.0 mW, 1.5 mW, and 2.0 mW are displayed in Table \ref{table:Model}. It can be observed that $L$ is proportional to the excitation power as expected with a proportionality factor of $\sim$10. Interestingly, $C$ also showed a power dependence. It decreased $\sim$5 mHz for each 0.5 mW increase in the laser power. This may be due to our temperature model failing to account for the differences that can occur between the used value of $T_R$ and the actual value of $T_R$ for higher laser powers. Since a higher laser power induces a greater local temperature change, it will have a greater effect on the surrounding temperature and may actually increase the effective reservoir temperature (temperature surrounding the emitters but far from our temperature probe). A higher effective reservoir temperature $T_R$ leads to a smaller temperature difference $T-T_R$, which leads to a smaller rate of conduction $C(T-T_R)$. Since our model treats $T_R$ as a fixed constant during fitting, the smaller rate of conduction is reflected in a smaller value of $C$ for higher powers.


\section{\label{sec:Conclusion}Conclusions}
In summary, we studied the effects of localized laser-induced heating on the PL spectrum of $\mathrm{V_{Si}}$ color centers in 4H-SiC. The effects of localized laser-heating were observable in our PL measurements for an excitation power of 0.5 mW (an intensity of 62.5 $\mathrm{W/mm^2}$ for a beam size of $\sim 8 $ \SI{}{\micro\meter\squared}) or higher at temperatures below 125 K. The laser-induced heating raised the local temperature of the emitters in the sample, leading to a decrease in the peak height, redshift of the center energy, and broadening of the linewidth in the PL of the V1, $\mathrm{V1^\prime}$, and V2 ZPLs arising from $\mathrm{V_{Si}}$ color centers in 4H-SiC. These effects are more pronounced for lower temperatures and higher powers. The emitters' local temperature and, hence, the temperature changes can be determined from the temperature dependence of the center energy of the ZPLs by using the Varshni equation. The largest local temperature change observed caused by the localized laser-induced heating was $\sim$25 K. It occurred when the sample was initially at a temperature of 5 K and was excited with a laser power of 2 mW, corresponding to an intensity of 250 W/$\mathrm{mm^{2}}$. The temperature of the emitters can be modeled as a system in contact with a thermal reservoir being heated by laser radiation. The model showed a good agreement with the data of the V1 and $\mathrm{V1^\prime}$ ZPLs for a large range of temperatures and powers and was able to predict a heating rate proportional to the laser power. 

The study shows that the effects due to laser-induced heating on the optical properties of $\mathrm{V_{Si}}$ color centers in 4H-SiC become significant with an excitation laser power as low as 0.5 mW when tightly focused. It is critical to account for laser heating effects in applications of 4H-SiC that employ laser excitation and emission from ZPLs of color centers. On the other hand, the sharp and bright ZPLs of color centers can facilitate the construction of integrated local temperature probes in 4H-SiC devices.



\section{\label{sec:Ack}Acknowledgement}
This work was supported by the Army Research Office (ARO) and was accomplished under Grant Number W911NF-23-1-0195. AM acknowledges the support by NSF via grant DMR-2122078. SR acknowledges the support by NASA award $\#$80NSSC19M0201.

\appendix

\section{Timed PL measurements of V2}\label{Apx:PL}
\begin{figure}[h]
    \centering
    \includegraphics[width=\columnwidth]{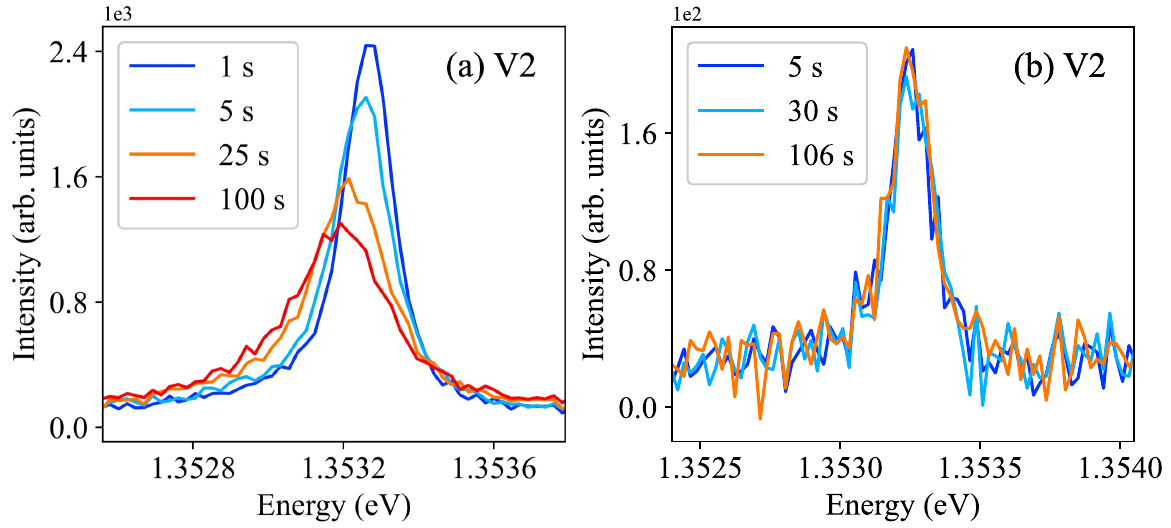}
    \caption{Timed PL measurements of the V2 ZPL conducted at a cryostat temperature of 10 K for laser powers of (a) 2 mW and (b) 0.03 mW. The legends indicate how long the laser illuminated the sample when the spectrum was taken.}
    \label{fig:DecayV2}
\end{figure}

Timed PL spectra of the V2 ZPL are shown in Fig. \ref{fig:DecayV2} for an excitation laser power of (a) 2 mW and (b) 0.03 mW. All measurements were taken at the cryostat temperature of 10 K. The legends in the figure indicate how long the laser illuminated the sample before the spectrum was taken. With a laser power of 2 mW, as shown in Fig. \ref{fig:DecayV2}(a), the spectra show a decrease in the peak height, redshift of the center energy, and broadening of the linewidth as the duration of laser illumination increases. With a laser power of 0.03 mW, no significant changes in the spectral characteristics of the V2 ZPL were observed, as shown in Fig. \ref{fig:DecayV2}(b). The timed PL measurements of the V2 ZPL show the same behavior as the V1 and V1$^\prime$ ZPLs.

\section{Lineshape fitting}\label{Apx:Fitting}
\begin{figure}[t]
    \centering
    \includegraphics[width=\columnwidth]{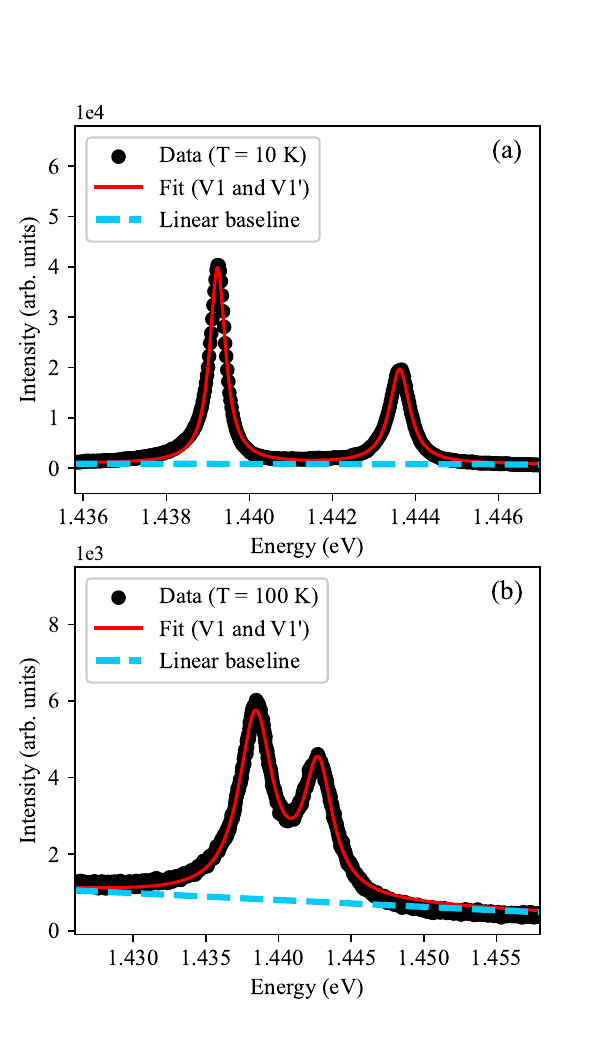}
    \caption{Fitting (red lines) of the PL spectra (black dots) of V1 and $\mathrm{V1^\prime}$ at cryostat temperatures of (a) 10 K and (b) 100 K. The linear baseline from the fitting is shown as a dashed blue line.}
    \label{fig:FittingV1}
\end{figure}

\begin{figure}[h]
    \centering
    \includegraphics[width=\columnwidth]{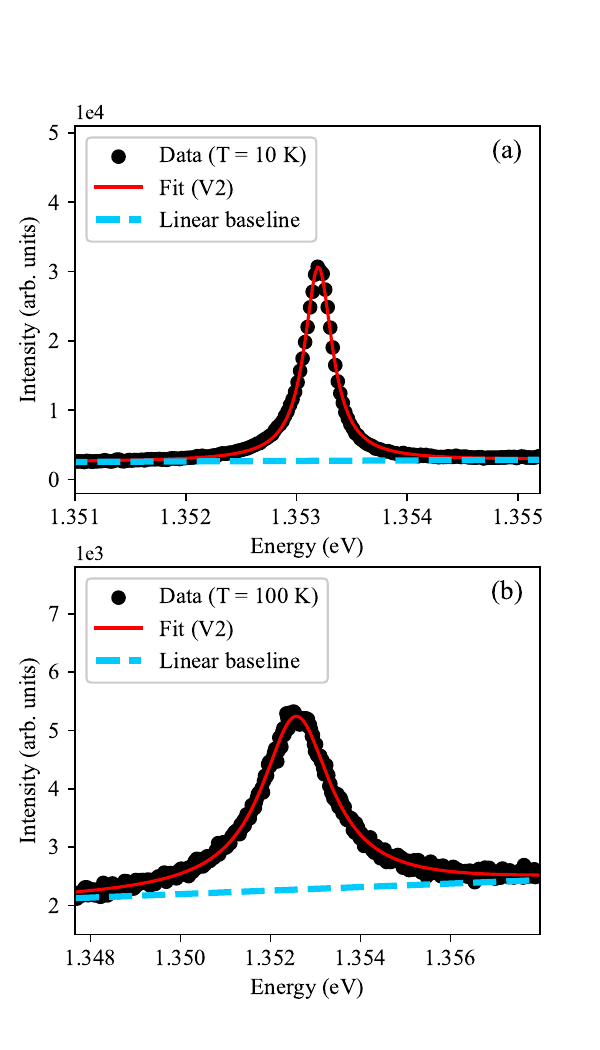}
    \caption{Fitting (red lines) of the PL spectra (black dots) of V2 at cryostat temperatures of (a) 10 K and (b) 100 K. The linear baseline from the fitting is shown as a dashed blue line.}
    \label{fig:FittingV2}
\end{figure}

The Lorentzian profile was used to fit the ZPL PL spectra to extract the peak height, center energy, and linewidth. A linear baseline was added to the Lorentzian profile to systematically account for the overlap between the wings of the PSB and the ZPL during the fitting. The V1 and $\mathrm{V1^\prime}$ ZPL PL spectra were fitted together using the following equation
\begin{equation}\label{eq:LorentzianV1}
     I = b + m*E 
     + \sum^2_{i=1} \frac{A_i\Gamma_i}{2\pi}\frac{1}{(E-E_i)^2 + (\frac{\Gamma_i}{2})^2},
\end{equation} 

\noindent where $I$ is the intensity, $b$ is the intercept of the linear baseline, $m$ is the slope of the linear baseline, $E$ is the energy, $A_i$ is the peak area, $\Gamma_i$ is the linewidth, and $E_i$ is the center energy of V1 when $i=1$ and of $\mathrm{V1^\prime}$ when $i=2$. The V2 ZPL PL spectra were similarly fitted but with one Lorentzian profile instead of two. Fig. \ref{fig:FittingV1} shows examples of the V1 and $\mathrm{V1^\prime}$ fitting for cryostat temperatures of (a) 10 K and (b) 100 K. Fig. \ref{fig:FittingV2} shows examples of the V2 fitting for cryostat temperatures of (a) 10 K and (b) 100 K. The fitting for both figures shows good agreement with the data. It can also be observed that the absolute value of the slope of the linear baseline (dashed blue line) increases with increasing cryostat temperature as the overlap between the wings of the PSB and the ZPLs increases.

From the fits, the peak height, $H$, can be extracted as a function of the peak area, $A$, and the linewidth, $\Gamma$, using the following equation
\begin{equation}\label{eq:Height}
    H = \frac{2A}{\pi\Gamma}.
\end{equation}

\section{Temperature dependence of low-power PL for V2}\label{Apx:LowP}

\begin{figure}[!]
    \centering
    \includegraphics[width=\columnwidth]{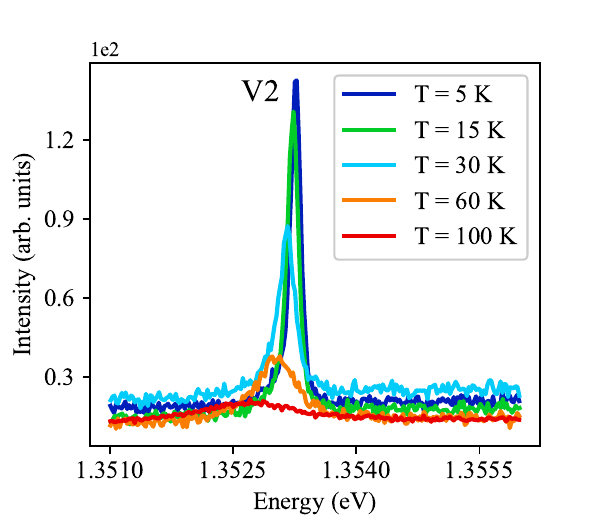}
    \caption{PL spectra taken with an excitation power of 0.03 mW at cryostat temperatures 5 K, 15 K, 30 K, 60 K, and 100 K. The V2 ZPL decreases in peak height, redshifts, and broadens with increasing temperature.}
    \label{fig:V2TempDep}
\end{figure}

Similar to the measurements done for the V1 and $\mathrm{V1^\prime}$ ZPLs in Section \ref{sec:LowP}, we performed a temperature dependence of the V2 ZPL PL using a power of 0.03 mW. A selection of the spectra from the temperature dependence can be observed in Fig. \ref{fig:V2TempDep}. The
legend of the figure indicates the cryostat temperature at which the spectrum was taken. A decrease in the peak height, redshift of the center energy, and broadening of the linewidth of the V2 ZPL with increasing temperature can be seen in the figure. This is the same behavior exhibited by the other ZPLs with temperature. It is also the same behavior as the one exhibited by the V2 ZPL over time when excited with high power (Fig. \ref{fig:DecayV2} (a)). 

\begin{figure}[h!tbp]
    \centering
    \includegraphics[width=\columnwidth]{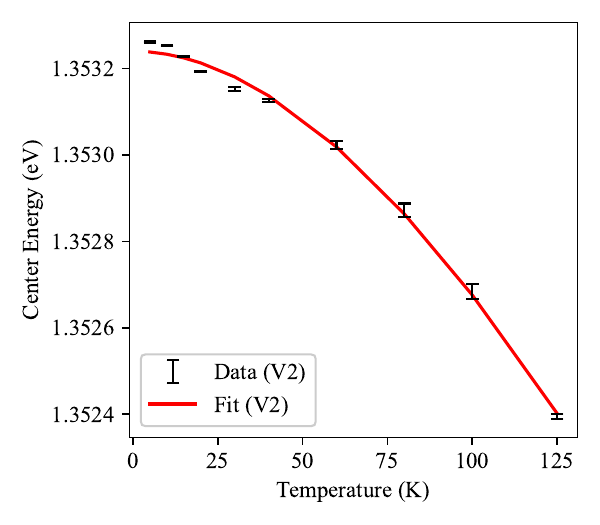}
    \caption{Center energy of V2 versus temperature measured for an excitation power of 0.03 mW. The Varshni equation was fitted to the data. The fit is displayed as the red line.}
    \label{fig:VarV2}
\end{figure}

 During the temperature dependence, three spectral measurements were taken at each temperature and fitted as described in Appendix \ref{Apx:Fitting} to obtain the center energy of V2. The center energy mean, $\bar{E}_0$, and the sample standard deviation, $S$, were calculated from the three measurements at each temperature$\bar{E}_0$ for V2 was graphed versus temperature in Fig. \ref{fig:VarV2} and the error bars of the figure were made from $S$. The relationship between the center energy of the V2 ZPL and temperature can be described by the Varshni equation, Eq. (\ref{eq:Varshni}), similarly to the V1 and $\mathrm{V1^\prime}$ ZPLs in Section \ref{sec:LowP}. The fit of the Varshni equation to the center energy of V2 versus temperature data is shown in Fig. \ref{fig:VarV2} as the red line. The fitting shows good agreement with the data as it did for V1 and $\mathrm{V1^\prime}$ in Section \ref{sec:LowP}. The Varshni fitting parameters for V2 are shown in Table \ref{table:Varshni}.


\newpage 

\bibliography{Bibliography}

\end{document}